# Coherent Modulation of the $YBa_2Cu_3O_{6+x}$ Atomic Structure by Displacive Stimulated Ionic Raman Scattering

R. Mankowsky[1,2], M. Först[1], T. Loew[3], J. Porras[3], B. Keimer[3] and A. Cavalleri[1,2,4]

[1] Max Planck Institute for the Structure and Dynamics of Matter, 22761 Hamburg, Germany.
[2] University of Hamburg, 22761 Hamburg, Germany.
[3] Max Planck Institute for Solid State Research, 70569 Stuttgart, Germany.
[4] Department of Physics, Oxford University, Clarendon Laboratory, Oxford OX1 3PU, UK

We discuss the mechanism of coherent phonon generation by Stimulated Ionic Raman Scattering, a process different from conventional excitation with near visible optical pulses. Ionic Raman scattering is driven by anharmonic coupling between a directly excited infrared-active phonon mode and other Raman modes. We experimentally study the response of $YBa_2Cu_3O_{6+x}$ to the resonant excitation of apical oxygen motions at 20 THz by mid-infrared pulses, which has been shown in the past to enhance the interlayer superconducting coupling. We find coherent oscillations of four totally symmetric ($A_g$) Raman modes and make a critical assessment of the role of these oscillatory motions in the enhancement of superconductivity.

Mid-infrared light pulses can be used to resonantly excite optically active lattice vibrations in solids to amplitudes as high as several per cent of interatomic distances. This technique has been shown to induce changes in the collective magnetic and electronic properties of many materials [1,2,3]. Integral to this optical control mechanism is the anharmonic coupling between the directly driven, optically active mode and other lattice vibrations [4,5,6]. The coupling is typically dominated by cubic anharmonicities and involves a displacive force that acts onto the crystal lattice in two ways.

Firstly, a phononic analogue to rectification in nonlinear optics causes a quasi-static displacement along the normal mode coordinates of all coupled Raman modes. Secondly, whenever the displacive force rises promptly compared to the period of any of the anharmonically coupled modes, coherent oscillatory motions of these modes are excited. This second effect is the stimulated equivalent of Ionic Raman scattering.

Here, we study the coherent optical response of $YBa_2Cu_3O_{6+x}$. Infrared-active apical oxygen motions are driven resonantly with mid-infrared pulses at 20 THz, under the conditions for which superconducting transport is transiently enhanced [7,8,9]. We find that for excitation with mid-infrared pulses of 140 fs duration, for which only modes with frequency <6 THz can be driven coherently, oscillations of four Raman modes are stimulated, involving displacements of the copper atoms along the crystallographic $c$ axis. This motion induces periodic changes in the in-plane O-Cu bond buckling and leads to an oscillatory transfer of charges between the $CuO_2$ planes and the Cu-O chains [9], effectively modifying the doping of the planes; an effect that may be part of the puzzle of optically enhanced superconductivity in this compound [7,8,9].

We next discuss the process of stimulated Ionic Raman scattering in more detail. The indirect excitation of coherent Raman modes by resonant excitation of large amplitude infrared-active (IR) vibrations in a solid is described to lowest order by the lattice Hamiltonian

$$H = \frac{1}{2}\omega_{IR}^2 Q_{IR}^2 + \frac{1}{2}\omega_R^2 Q_R^2 - a_{12} Q_{IR} Q_R^2 - a_{21} Q_{IR}^2 Q_R, \tag{1}$$

where $(\omega_{IR}, Q_{IR})$ and $(\omega_R, Q_R)$ denote the respective frequency and normal coordinates of the directly excited IR mode and of any anharmonically coupled mode. In this equation, $a_{12}$ and $a_{21}$ are the coupling constants. For centrosymmetric crystals like $YBa_2Cu_3O_{6+x}$, the term $a_{12} Q_{IR} Q_R^2$ is forbidden as any infrared mode $Q_{IR}$ is odd (breaks inversion symmetry), whereas $Q_R^2$ is even (conserves inversion symmetry). Furthermore, $Q_R$ in $a_{21}Q_{IR}^2Q_R$ must be a Raman mode as $Q_{IR}^2$ is even. The Hamiltonian thus reduces to

$$H = \frac{1}{2}\omega_{IR}^2 Q_{IR}^2 + \frac{1}{2}\omega_R^2 Q_R^2 - a_{21}Q_{IR}^2 Q_R. \tag{2}$$

The corresponding dynamical response of the modes is described by the equations of motion

$$\ddot{Q}_{IR} + 2\gamma_{IR}\dot{Q}_{IR} + \omega_{IR}^2 Q_{IR} = f(t) + 2a_{21}Q_{IR}Q_R, \tag{3}$$

$$\ddot{Q}_R + 2\gamma_R\dot{Q}_R + \omega_R^2 Q_R = a_{21}Q_{IR}^2. \tag{4}$$

Dissipation is accounted for by the terms containing $\gamma$, which is the inverse lifetime of the respective phonon mode $\gamma = \tau^{-1}$. The equation for the IR mode $Q_{IR}$ describes a damped harmonic oscillator driven by the electric field of the mid-infrared pulse $f(t) = A(t)e^{i\omega_{IR}t}$, with $A(t)$ being the Gaussian envelope of the pulse. Upon excitation, the atoms perform oscillations along the IR mode eigenvector about their equilibrium positions as shown in Fig. 1a in light red. This motion, for a finite anharmonic coefficient $a_{21}$, exerts a directional force $F(t) = a_{21}Q_{IR}^2(t)$ proportional to $Q_{IR}^2$ onto the coupled Raman mode (Fig. 1b).

Hence, the atoms experience a displacement along *all* coupled Raman-mode eigenvectors. This displacement may occur fast or slow compared to the eigenfrequency of each coupled mode. It relaxes back to equilibrium over a timescale that is one half of the dephasing time of the *infrared* mode $\tau_{IR} = 1/\gamma_{IR}$ assuming no transition into a metastable state takes place. The one half factor for the relaxation time descends from the fact that the squared amplitude of the IR mode appears in the driving term in Equation (4). This effect has been studied in detail in Refs. [6] and [9].

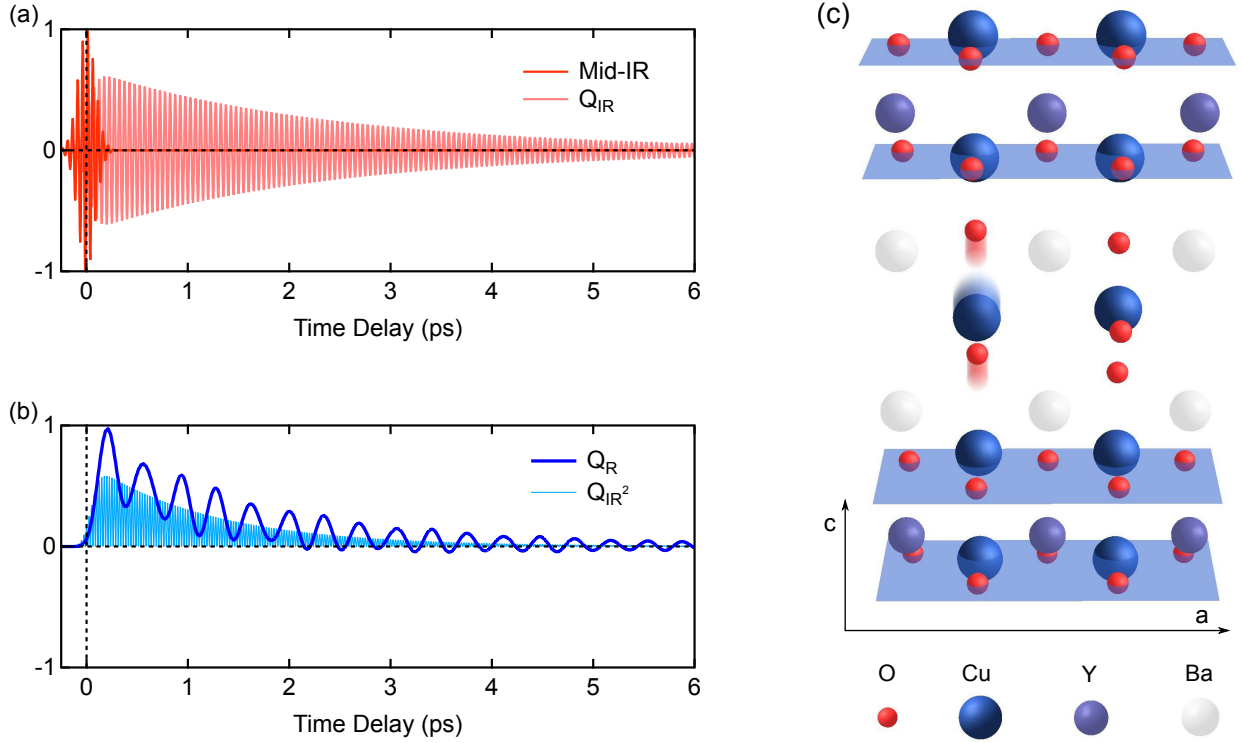

FIG. 1. (a) Response of the infrared-active normal coordinate $Q_{IR}$ (light red) to resonant excitation by a 140fs mid-infrared pulse (red). (b) Within cubic coupling, a directional force proportional to $Q_{IR}^2$ (light blue) is exerted on the coupled mode $Q_R$, which displaces the atoms as long as the infrared mode oscillates coherently and drives coherent oscillations for modes $Q_R$ with a period long compared to the rise time of this force. (c) YBa$_2$Cu$_3$O$_{6+x}$ is a layered superconductor in which CuO$_2$ planes, shown in blue, form bilayers separated by an insulating Yttrium layer. Holes are doped in these planes by changing the oxygen content of Cu-O chains, which are alternatingly filled and empty in the ortho-II ordered structure of YBa$_2$Cu$_3$O$_{6+x}$ for x=0.5 or 0.55. The resonantly excited B$_{1u}$ mode consists of movements of the apical oxygen atoms between these bilayers at chain sites with an oxygen vacancy.

Secondly, as discussed in the introduction, the coupled Equations (3) and (4) predict that all of the displaced modes, which have a long enough eigenperiod, will exhibit coherent oscillations about the displaced atomic positions. Specifically, this happens only for those modes with eigenperiod $T_R$ long compared to the rise time of the directional force, or, equivalently, compared to the width of the mid-infrared pulse envelope $\Delta_{IR}$ driving the odd mode $Q_{IR}$. The precise oscillation amplitude will further depend on the anharmonic coupling constant but also on the rise and decay times of the driving force $F(t)$ compared to the eigenperiod of the Raman mode. This is again captured by the two classical equations of motion. Further, unlike for the displacive response discussed above, these oscillations relax over a timescale that is determined by the lifetime of the Raman mode $\tau_R = 1/\gamma_R$, which may be far longer than $\tau_{IR}/2$ (see blue and red oscillations in Figure 1).

Finally, the phase of the oscillations may be zero or finite, depending on whether the process is in the *impulsive* or *displacive* limit. In the impulsive limit, ($T_R \gg \Delta_{IR}$, $\tau_{IR}$ and $\omega_{IR} \gg \omega_R$) the infrared-active mode decays back to its ground state before the Raman mode has started oscillating. Hence, Raman oscillations take place about the equilibrium lattice coordinates, with $Q_R(t) \sim \sin(\omega_R t)$. In the displacive limit, ($T_R \gg \Delta_{IR}$, $T_R \ll \tau_{IR}$ and $\omega_{IR} \gg \omega_R$) the IR mode and the displacive response decay slowly, and the Raman excitation occurs about the displaced positions. The force $F(t)$ acting on the Raman mode can be approximated by a step function and $Q_R(t) \sim (1-\cos(\omega_R t))$. In most cases, the oscillations will have a phase that is neither *sine* nor *cosine*, or the phase of the Raman mode may change over time [10].

For conventional electronic stimulated Raman scattering, the equation of motion (Eq. (4)) has the same structure and the same limits discussed above can be derived [11,12]. The physical process is however fundamentally different, as in conventional stimulated Raman scattering the driving force derives from electronic transitions [13]. On the contrary, in ionic stimulated Raman scattering the driving force solely depends on lattice variables [14,15,16,17]. The excitation process is therefore most sensitive to phase transitions that affect the lattice.

In the following, we discuss experiments performed on $YBa_2Cu_3O_{6.5}$ and $YBa_2Cu_3O_{6.55}$, two bilayer high-temperature superconductors with a critical temperature of $T_c = 50$ K and $T_c = 61$ K, respectively. These two compounds crystallize in a centrosymmetric orthorhombic structure with $D_{2h}$ symmetry. The bilayers comprise two conducting $CuO_2$ layers in the *ab* plane, which are separated by an insulating Yttrium layer (Fig. 1c). The $CuO_2$ planes are hole doped by adding oxygen atoms to the Cu-O chains along the *b* axis, which are vacant of oxygen atoms in the parent compound $YBa_2Cu_3O_6$. The samples measured in this work exhibited ortho-II ordering of the oxygen atoms, corresponding to a structure for which alternate Cu-O chains are filled and empty. The $YBa_2Cu_3O_{6.5}$ sample exhibited only short-range ordered domains, whereas the $YBa_2Cu_3O_{6.55}$ samples showed long-range ordering of these chains [18].

In samples with these doping levels, the resonantly excited $B_{1u}$ symmetry mode at 20 THz consists of movements of the apical oxygen atoms between bilayers at vacant chain sites only (blurred motion in Fig 1c) [19]. The long range ordering of the chain vacancies in the $YBa_2Cu_3O_{6.55}$ sample might therefore influence the structural dynamics. According to the symmetry argument discussed above, nonlinear coupling is restricted to modes of $A_g$ symmetry as the product group of $B_{1u}$ with itself is $A_g$. We thus expect a transient displacement of the crystal lattice along all of these modes with finite anharmonic coupling. Because atomic motions along Raman coordinates modulate the polarizability tensor, these motions become observable as changes in the reflectivity of the material. We used 140 fs mid-infrared pulses (15μm, 10% bandwidth) with a fluence of 2.5 mJ/cm$^2$ to drive the sample into its transient state, and probed the reflection of 35 fs pulses at 800 nm wavelength. Under this condition, we expect the excitation of modes up to ~6 THz in frequency. According to their Raman tensor for the orthorhombic $D_{2h}$ point group, the $A_g$ Raman modes are observable but have different tensor elements for probe polarizations in-plane along *a* and *b*, as well as out-of plane along the *c* direction [20].

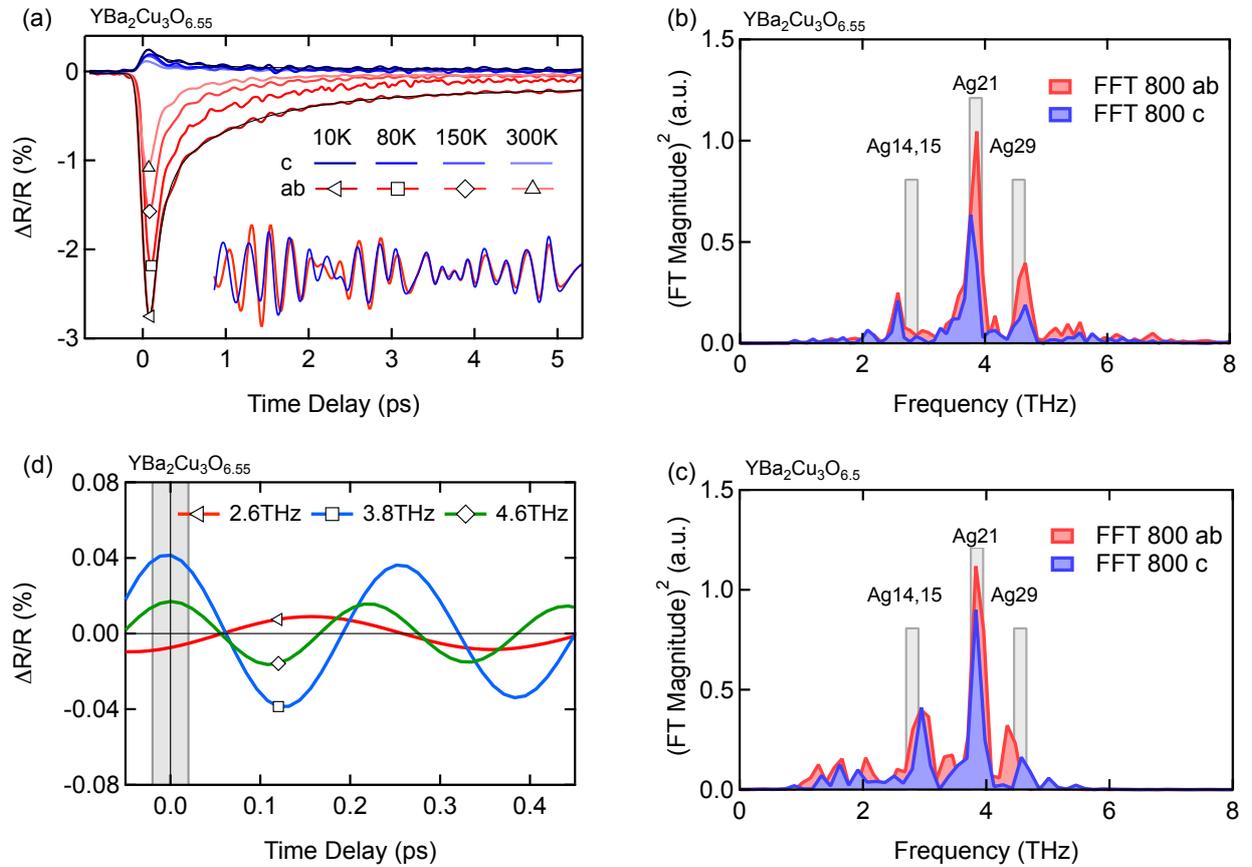

FIG. 2. (a) Time-resolved reflectivity changes at 800 nm following resonant excitation of the B1u infrared mode in YBa$_2$Cu$_3$O$_{6.55}$. Changes for probe polarizations out of plane and in plane are shown in blue and red, respectively. The inset shows the oscillatory signal at 10 K sample temperature, obtained by subtracting a fit to the data (thin black line). (b) The Fourier transformations of these oscillations, showing spectral weight at four frequencies coinciding with Ag phonon modes. (c) Same as panel (b) for YBa$_2$Cu$_3$O$_{6.5}$. (d) Zoom into the individual oscillatory components in YBa$_2$Cu$_3$O$_{6.55}$ (10 K, in plane), extracted from fitting the oscillations shown in panel (a). The phase of the oscillatory components at time zero is cosinelike.

The experimental results, reported for both out-of-plane and in-plane polarized probe pulses, are shown in Figure 2a for YBa$_2$Cu$_3$O$_{6.55}$. The oscillatory response was obtained by subtracting a fit to the data consisting of an error function and a triple-exponential decay (thin black line) and is shown in the inset for 10 K sample temperature. While sign and size of the total response are different for the two orthogonal polarizations, we find no difference between the amplitudes and absolute phases of the oscillations.

For $YBa_2Cu_3O_{6.55}$ and $YBa_2Cu_3O_{6.5}$, we find three dominant frequency components (Fig. 2b,c), which can be attributed to four $A_g$ phonon modes ($A_g14$, $A_g15$, $A_g21$ and $A_g29$) [9] that are shown in Figure 3a). The numbers denote the index of the respective phonon modes, as sorted according to their frequencies. We find no significant differences in the response to the excitation between the two samples. We stress that these modes could not be detected in the x-ray experiments of Ref. [9], as the signal to noise ratio was not sufficiently high.

Note that these oscillatory modes were also observed in $YBa_2Cu_3O_{6.9}$ after excitation with 2eV pulses [21,22,23]. In these experiments, large changes were found in the relative amplitudes of these modes and relaxation dynamics when crossing the critical temperature $T_c$, as the excitation mechanism relied on optical transitions between electronic states that strongly change through the superconducting transition. Here, we only find a continuous increase in both, phonon amplitudes and non-oscillatory components upon decreasing temperature, as only small changes in the equilibrium infrared phonon spectrum take place at $T_c$.

The observed oscillations at 2.6THz, 3.8THz and 4.6THz in $YBa_2Cu_3O_{6.55}$ are shown in Figure 2d, displaying a clear cosine phase at time zero (black line, grey area is the uncertainty), indicative of displacive rather than impulsive mechanism ($T_R \ll \tau_{IR}$). This relation can be validated by estimating the relaxation time of the IR mode $\tau_{IR}$ from the decay time of the atomic displacements $\tau_D$, which has been measured by x-ray diffraction and reported in Ref. [9], by $\tau_{IR} = 2\tau_D$. We obtain a relaxation time $\tau_{IR} = 2.4 ps$, which is larger by factors of 11, 9 and 6 compared to the periods of the observed oscillations at 2.6THz, 3.8THz and 4.6THz, respectively, consistent with a cosine phase at time zero. We can attribute these oscillations to the four lowest frequency $A_g$ modes of the ortho-II ordered $YBa_2Cu_3O_{6+x}$ structure. From the presented data alone, the amplitudes of the atomic motions cannot be quantified, as the changes in the electronic polarizability at 800 nm may be different for each mode and are not known here [13].

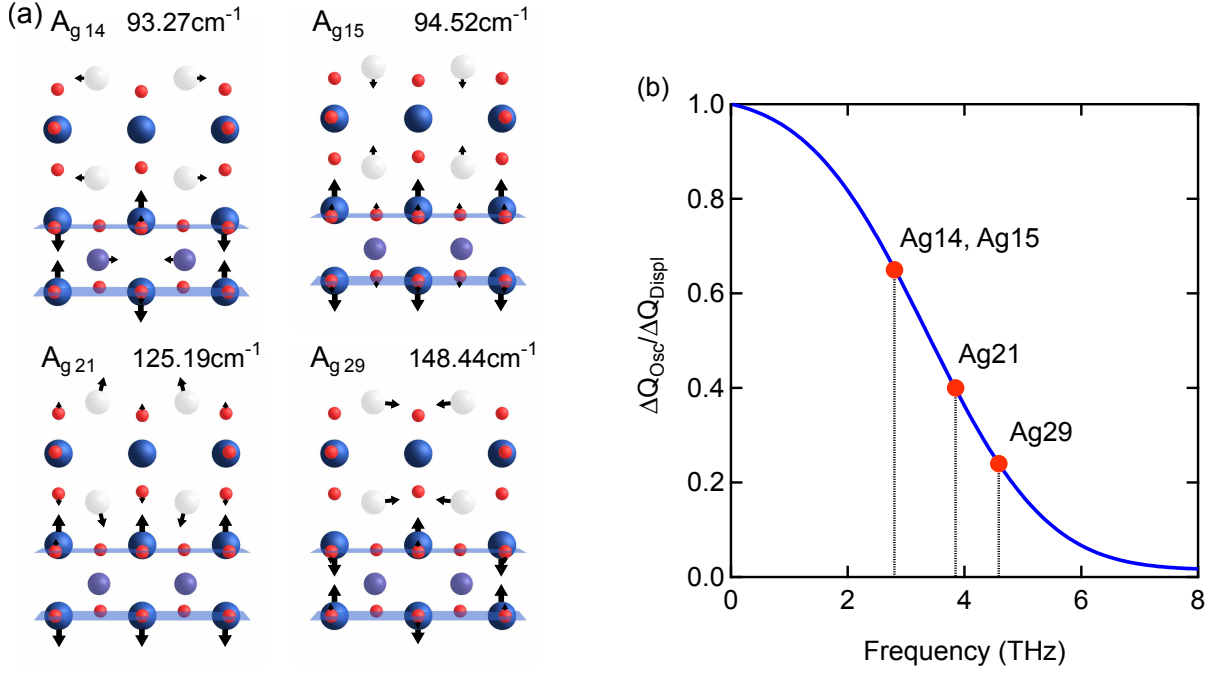

FIG. 3. (a) Four lowest frequency $A_g$ modes of the ortho-II ordered YBa$_2$Cu$_3$O$_{6+x}$ structure. (b) Calculated ratio between oscillation and displacement amplitude of these modes after excitation of the $B_{1u}$ infrared mode with 140fs mid-infrared pulses.

|  | $\Delta d_{1,Displ}$ (pm) | $\Delta d_{2,Displ}$ (pm) | Frequency (THz) | $\Delta Q_{Osc} / \Delta Q_{Displ}$ |
|---|---|---|---|---|
| $A_g 14$ | 0.29 | -0.30 | 2.80 | 0.64 |
| $A_g 15$ | 0.81 | 0.76 | 2.83 | 0.64 |
| $A_g 21$ | 0.73 | 0.63 | 3.85 | 0.40 |
| $A_g 29$ | 0.24 | -0.34 | 4.59 | 0.24 |

TABLE 1. Numbers used for the estimation of the real-space amplitudes of the $A_g$ modes oscillatory motion. The frequencies and displacement amplitudes $\Delta d_{Displ}$ of the $A_g$ modes are taken from the DFT calculations for YBa$_2$Cu$_3$O$_{6.5}$ described in Ref. [9]. The relative amplitudes of oscillatory and displacive components, $\Delta Q_{Osc} / \Delta Q_{Displ}$ are calculated from Equations (3) and (4) for 140fs pump pulses.

We obtain a qualitative estimate of the real-space amplitudes by starting from the measurements of the underlying rectified displacement by ultrafast hard x-ray diffraction of Ref. [9]. For the 140 fs mid-infrared pulses used to drive the odd apical oxygen mode, the relative amplitude of oscillatory and displacive responses only depends on the Raman mode frequency and can be calculated from Equations (3) and (4) (Fig. 3b). From these calculations and the displacement amplitudes of the four $A_g$ modes, we estimate the oscillatory amplitudes of the respective vibrations for the same fluence of Ref. [7,8,9] of 4mJ/cm$^2$ (see Table 1). The atomic motions of these modes (Fig. 3a) are dominated by a change in distance between Cu atoms of neighboring $CuO_2$ planes along the crystallographic c-axis (Fig. 4a). We estimate oscillation amplitudes in these distances of ~0.9pm and ~0.5pm at vacant ($d_1$) and filled ($d_2$) chain sites, relaxing with a decay time of 3 ps, as shown in Figure 4b.

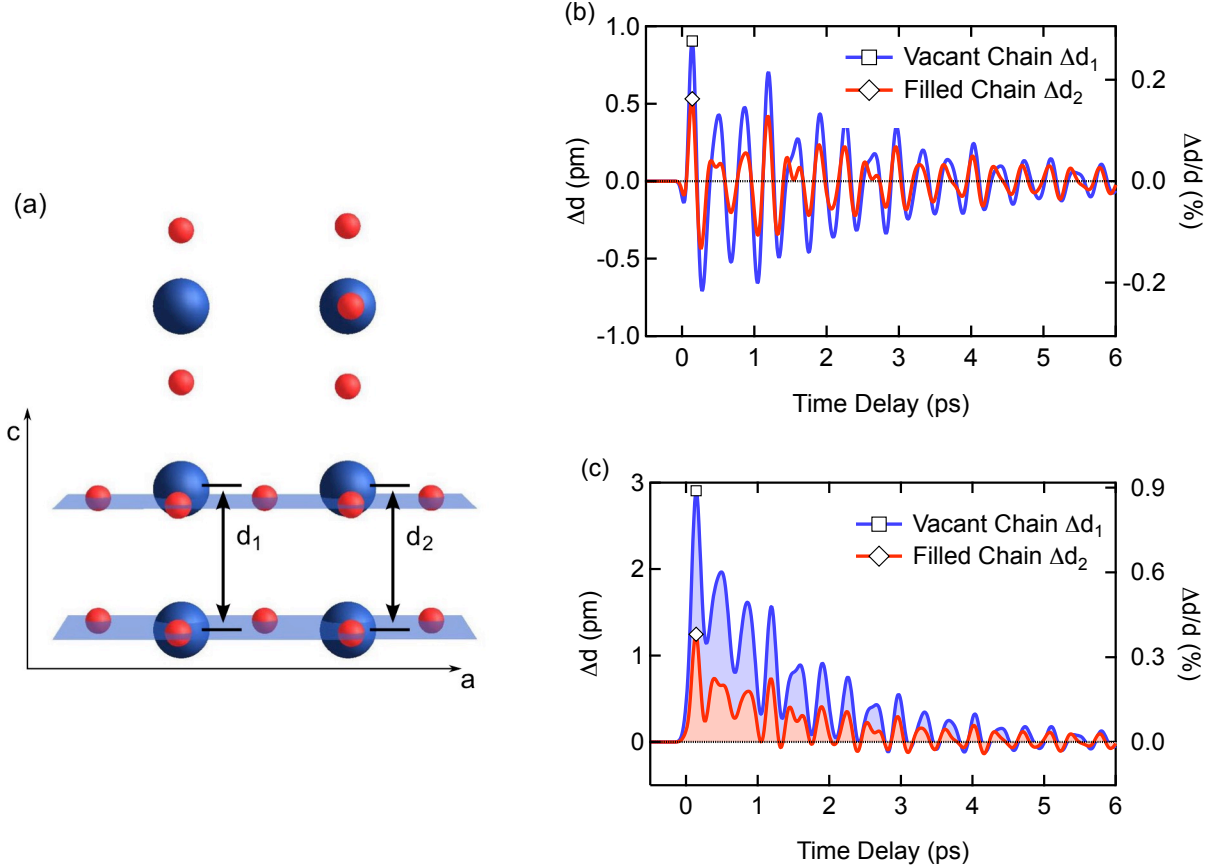

FIG. 4. (a) The combined atomic motions of the $A_g$ modes primarily involve c-axis movement of the planar Cu atoms. To describe the structural dynamics, we define $d_1$ and $d_2$ as the distances between Cu atoms (blue) of neighboring $CuO_2$ planes at vacant and filled chain sites, respectively. (b) Changes in the distances $d_1$ and $d_2$ due to excitation of the $A_g$ vibrations. We estimate oscillation amplitudes of 0.9pm and 0.5pm in $d_1$ and $d_2$, with a relaxation time of 3 ps. (c) The full dynamics including the displacive response involve an increase in both $d_1$ and $d_2$ accompanied by oscillations.

At this stage we can speculate if the phenomenon of enhanced interlayer coupling, sometimes referred to as light-induced superconductivity, may or may not be connected to these oscillations. According to the calculations of Ref. [9], the oscillatory motions reported above will couple to the density of states of the $d_z^2$ Cu orbitals in the chains and the planar Cu $d_{x^2-y^2}$ orbitals, and induce a charge transfer between plane and chain. We speculate here that these oscillations will modulate the interlayer tunneling and with it the coupling between the planes. The full dynamics shown in Figure 4c, which includes the displacive response, decreases the distances between plane and chain, and forces a net charge transfer from the planar Cu to the chain Cu, effectively increasing the hole doping of the planes. This charge transfer is emerging as a key process accompanying the superconducting transition, as shown, for instance, by the self-doping effect recently found to accompany the temperature-driven metal-superconductor state in $YBa_2Cu_3O_{6.9}$ [24].

In a recent theoretical paper [25], we showed how modulation of the interlayer couplings may cause a reduction and parametric cooling of phase fluctuations, an effect that may aid superconductivity in the driven state. In Ref. [25], we found that this effect is strongest for modulations at the difference frequency between the intra-bilayer and inter-bilayer Josephson plasmon, which in $YBa_2Cu_3O_{6.5}$ are at ~13 THz and ~1 THz, respectively. The modulations detected here are strongest at ~4 THz, and cannot account for such parametric cooling directly. For this effect to be effective the coupling would have to occur at the third harmonic frequency of the modulation, an effect that is unlikely but not impossible. Yet, we cannot exclude that modes may cause dynamical stabilization of interlayer fluctuations by other, related mechanisms different from the parametric cooling of Ref. [25], which may involve modulations of the electronic properties [26].

In summary, we have studied coherent phonon generation by displacive stimulated ionic Raman scattering in $YBa_2Cu_3O_{6.5}$ and $YBa_2Cu_3O_{6.55}$. We have reported measurements of coherent oscillations of four Ag phonon modes, triggered by anharmonically coupled apical oxygen motions, directly driven by mid-infrared pulses. This excitation is significant because in other experiments it has been shown to enhance superconducting interlayer tunneling. We present a model describing the generation mechanism based on cubic coupling of a directly driven infrared mode to other Raman modes. We combine this model with results of previous ultrafast x-ray diffraction experiments to estimate the oscillatory atomic motions and their amplitudes. We propose that the motions modulate the interlayer tunneling between adjacent $CuO_2$ planes, which may contribute to promoting superconductivity by periodic modulation of the Hamiltonian parameter.

Acknowledgements

We thank Alaska Subedi for providing the results from Density Functional Theory calculations used for the quantitative evaluation of the oscillation amplitudes (Ref. [9]).

The research leading to these results has received funding from the European Research Council under the European Union's Seventh Framework Programme (FP7/2007-2013) / ERC Grant Agreement n° 319286 (QMAC).

[1] M. Rini, R. Tobey, N. Dean, J. Itatani, Y. Tomioka, Y. Tokura, R. W. Schoenlein, and A. Cavalleri, Nature (London) **449**, 72 (2007).

[2] M. Först, R. I. Tobey, S. Wall, H. Bromberger, V. Khanna, A. L. Cavalieri, Y.-D. Chuang, W. S. Lee, R. Moore, W. F. Schlotter, J. J. Turner, O. Krupin, M. Trigo, H. Zheng, J. F. Mitchell, S. S. Dhesi, J. P. Hill, and A. Cavalleri, Phys. Rev. B **84**, 241104 (2011).

[3] A.D. Caviglia, R. Scherwitzl, P. Popovich, W. Hu, H. Bromberger, R. Singla, M. Mitrano, M.C. Hoffmann, S. Kaiser, P. Zubko, S. Gariglio, J.-M. Triscone, M. Först, and A. Cavalleri, Phys. Rev. Lett. **108**, 136801 (2012)

[4] M. Först, C. Manzoni, S. Kaiser, Y. Tomioka, Y. Tokura, R. Merlin, and A. Cavalleri, Nature Phys. **7**, 854 (2011).

[5] M. Först, R. Mankowsky, H. Bromberger, D.M. Fritz, H. Lemke, D. Zhu, M. Chollet, Y. Tomioka, Y. Tokura, R. Merlin, J.P. Hill, S.L. Johnson, and A. Cavalleri, Solid State Commun. **169**, 24 (2013)

[6] A. Subedi, A. Cavalleri, and A. Georges, Phys. Rev. B **89**, 220301 (2014)

[7] S. Kaiser, C.R. Hunt, D. Nicoletti, W. Hu, I. Gierz, H.Y. Liu, M. Le Tacon, T. Loew, D. Haug, B. Keimer, and A. Cavalleri, Phys. Rev. B **89**, 184516 (2014)

[8] W. Hu, S. Kaiser, D. Nicoletti, C. R. Hunt, I. Gierz, M. C. Hoffmann, M. Le Tacon, T. Loew, B. Keimer, and A. Cavalleri, Nature Mat. **13**, 705 (2014)

[9] R. Mankowsky, A. Subedi, M. Först, S.O. Mariager, M. Chollet, H. T. Lemke, J. S. Robinson, J. M. Glownia, M. P. Minitti, A. Frano, M. Fechner, N. A. Spaldin, T. Loew, B. Keimer, A. Georges, and A. Cavalleri, Nature **516**, 71 (2014)

[10] M.F. DeCamp, D.A. Reis, P.H. Bucksbaum, and R. Merlin, Phys. Rev. B **64**, 092301 (2001)

[11] G.A. Garrett, T.F. Albrecht, J.F. Whitaker, and R. Merlin, Phys. Rev. Lett. **77**, 3661 (1996)

[12] R. Merlin, Solid State Commun. **102**, 207-220, (1997)

[13] T.E. Stevens, J. Kuhl, and R. Merlin, Phys. Rev. B **65**, 144304 (2002)

[14] A.A. Maradudin, and R.F. Wallis, Phys. Rev. B **2,** 4294 (1970)

[15] R.F. Wallis, and A.A. Maradudin, Phys. Rev. B **3,** 2063 (1971)

[16] T.P. Martin, and L. Genzel, phys. stat. sol. (b) **61,** 493 (1974)

[17] D.L. Mills, Phys. Rev. B **35**, 9278 (1987)

[18] S. Blanco-Canosa, A. Frano, E. Schierle, J. Porras, T. Loew, M. Minola, M. Bluschke, E. Weschke, B. Keimer, and M. Le Tacon, Phys. Rev. B, **90**, 054513 (2014)

[19] C.C. Homes, T. Timusk, D.A. Bonn, R. Liang, and W.N. Hardy, Physica C **254,** 265-280 (1995). (1995).

[20] W. Hayes, and R. Loudon, Scattering of Light by Crystals, Chap. 1., Wiley, New York, (1978)

[21] W. Albrecht, Th. Kruse, and H. Kurz, Phys. Rev. Lett. **69**, 1451 (1992)

[22] W.A. Kutt, W. Albrecht, and H. Kurz, IEEE J. Quantum Electron. **28**, 2434 (1992)

[23] I.I. Mazin, A. I. Liechtenstein, O. Jepsen, O.K. Andersen, and C.O. Rodriguez, Phys. Rev. B **49**, 9210 (1994)

[24] M. Magnuson, T. Schmitt, V.N. Strocov, J. Schlappa, A. S. Kalabukhov, and L. C. Duda, Scientific Reports **4**, 7017 (2014)

[25] S. Denny, S. Clark, Y. Laplace, A. Cavalleri, and D. Jaksch, ArXiv: 1411.3258

[26] R. Höppner, B. Zhu, T. Rexin, A. Cavalleri, and L. Mathey, Phys. Rev. B **91**, 104507 (2015).